\documentclass[jkps,preprint,showpacs,showkeys]{revtex4}
\usepackage{graphicx}
\usepackage{amssymb}
\usepackage{bm}

\begin{document}
\setcounter{page}{1}
\title[]{Relativistic Spin Operator and Lorentz Transformation of the Spin State of a Massive Dirac Particle}
\author{Taeseung \surname{Choi}}
\email{tschoi@swu.ac.kr}
\thanks{Fax: +82-2-970-5907}
\affiliation{Division of General Education, Seoul Women's University, Seoul 139-774,}
\affiliation{School of Computational Sciences, Korea Institute for Advanced Study, Seoul 130-012, Korea }
\date[]{}

\renewcommand{\abstractname}{}
\begin{abstract}
  We have shown that the covariant relativistic spin operator is equivalent to the spin operator commuting with the 
  free Dirac Hamiltonian. This implies that the covariant relativistic spin operator is a good quantum 
  observable. The covariant relativistic spin operator has a pure quantum contribution that does not exist in the classical 
  covariant spin operator. Based on this equivalence, reduced spin states can be clearly defined.
  We have shown that depending on the relative motion of an observer, the change in the entropy of a reduced spin 
  density matrix sweeps through the whole range.
\end{abstract}

%Use showkeys class option if keyword
\maketitle
%\maketitle

{\centering\section{Introduction}}

The spin of a massive Dirac particle is $1/2$, so it can be used as a qubit under the condition that a detector is ideal
in the sense that it does not respond to the momentum of the particle. The statistical prediction of measurements 
for a spin is provided by the reduced spin density matrix, which can be obtained by partial tracing  
over momentum degrees of freedom.
 The interesting fact is that the purity of the spin, which is described by the
Von Neumann entropy of the reduced spin density matrix, changes under the Lorentz transformation \cite{Peres}.
This would be expected because the spin of a massive Dirac particle rotates depending on the momentum of the particle 
under the general Lorentz transformation as Wigner has noted \cite{Wigner}.
By the Wigner rotation, the entanglement of spin and momentum degrees of freedom can be changed in general.

The seminal work by Peres {\it et al.} \cite{Peres}, which noticed the relativistic depurification of the spin 
by using two-component spinors, has not been fully appreciated. 
There were arguments whether the reduced spin density
matrix in Ref. 1 is a proper one \cite{Peres1}. 
This is partly because the physical spin observable was not noted in that paper \cite{Peres}. 
The change of the spin entropy can be described by the reduced spin density matrix which does not require 
the explicit form of a spin observable.
For the measurement of the spin entropy, however, the physical spin observables must be specified.  
Because of the lack of knowledge about the relativistic spin, most later works, which have developed the new field known 
as the relativistic quantum information \cite{Czachor, Alsing, Ahn, Czachor1, Peres2, Terashima,  
Lee, Kim, Caban1, Caban2, Friis, Gingrich, Vedral, Choi}, 
have been done by using spins different from the one implicitly given in the paper \cite{Peres}.

Gingrich and Adami have used the 4-dimensional Dirac eigen-spinors, which are the spin 
part of the solution for the covariant Dirac equation, because the eigen-spinors show manifest Lorentz covariance \cite{Gingrich}.
The Dirac eigen-spinor in a laboratory frame, where a particle is moving, 
has momentum-dependent components, so partial tracing over the momentum was criticized \cite{Czachor1}.
The momentum dependence in the components of the Dirac eigen-spinors can be eliminated in the special representation 
given by Foldy and Woutheysen \cite{Foldy}. 
We call this the Foldy-Woutheysen (FW) representation.
Therefore, in the FW representation, the meaning of partial tracing over the momentum is clear so that the reduced spin 
density matrix can be well defined.
The two spin density matrices for the same positive energy particle, one in the FW representation and 
the other in the covariant Dirac representation, give the same entropy. 
In this sense the results given by Peres {\it et. al} \cite{Peres} and Gingrich and Adami \cite{Gingrich}
 can be justified. 
 Recently it was noticed, however, that the reduced spin density matrix was meaningless because 
the spin of the relativistic particle could not be measured independently of its momentum \cite{Vedral}. 
This conclusion is erroneous because the classical spin operator, not 
the full quantum spin operator, had been considered \cite{ChoiNote}.

In the FW representation, there is a mean spin operator that commutes with the free Dirac Hamiltonian so that 
it is a good quantum observable. 
We call this spin operator the FW mean spin operator. G$\ddot{u}$rsey and Ryder \cite{Gursey} have shown that the covariant 
relativistic spin operator for positive energy states, 
which is obtained by using the Lorentz transformation in the spinor representation, is
the same as a FW mean spin operator for positive energy states. 
The equivalence for negative energy states, however, is not clear because the forms of two operators are different.
In this paper, we will show explicitly the equivalence between the covariant relativistic spin operator and 
the FW mean spin operator for the whole of spinor space. 
 The difference of the covariant relativistic spin operator from the classical covariant spin operator will also be discussed. 
 This will clarify the long-standing controversy regarding what are the proper spin operator
 and the reduced spin entropy in the laboratory frame.  
We will also study the change of the relativistic spin entropy    
in the FW representation in which the tracing over the momentum has a clear meaning.
The results show that the pure spin state can become a completely-mixed spin state and vice versa. 
 In Sections \ref{sec:CRS} and \ref{sec:CAP}, we review the covariant relativistic spin operator and 
the FW mean spin for clarity and self-containedness.
In Section \ref{sec:Equ}, we will show the equivalence of the two spin operators. 
In Section \ref{sec:RSD}, we will discuss the change of reduced spin density matrices
under Lorentz transformation.
In Section \ref{sec:SUM}, we will summarize our results. \\

{\centering\section{covariant relativistic spin}\label{sec:CRS}}

The spin operator in the laboratory frame can be obtained from the spin operator
in the particle rest frame by using the Lorentz transformation . 
This spin operator is a good candidate for a relativistic spin operator, and 
we call this spin operator the covariant relativistic spin operator.
We will review the procedure to obtain the covariant relativistic spin in a fully covariant way and discuss 
its difference from the classical covariant spin used to define the covariant spin magnetic dipole 
moment. For completeness, we also review another approach \cite{Ryder99} to obtain the covariant relativistic spin
operator by using Pauli-Lubanski vector.\\

{\raggedright\subsection{Lorentz Covariant Approach}}
Let us consider the following covariant Dirac equation for a free massive Dirac particle
in coordinate space:
\begin{eqnarray}
\label{eq:Dirac}
(i\gamma^\mu \partial_\mu -m) \psi(p)=0,
\end{eqnarray}
where
\begin{eqnarray}
\gamma^0 = \left(
\begin{array}{cc} I & 0 \\
0 & -I \end{array} \right),~
\gamma^i=  \left( \begin{array}{cc} 0 & \sigma_i \\ -\sigma_i & 0 \end{array} \right) \mbox{ for } i=1,2,3,
\end{eqnarray}
and $\sigma_i$ are the usual Pauli matrices. $\gamma^\mu$ are Dirac matrices. 
We use the sign convention $\eta_{\mu\nu}=\mbox{diag}(+,-,-,-)$ for the metric and
the natural units $c=\hbar=1$. The summation conventions are also used. The Greek index $\mu$ runs $0,1,2,3$ and 
the Latin index $i$ runs $1,2,3$.
The covariant Dirac equation has two positive energy solutions, $u(p,s)e^{-ip_\mu x^\mu}$, 
and two negative energy solutions, $v(p,s)e^{ip_\mu x^\mu}$, where $s=\pm 1$. 
$e^{-ip_\mu x^\mu}$ is the coordinate representation of the momentum eigenstate. 
$u(p,s)$ and $v(p,s)$ are the spin parts of the Dirac solutions and are called the positive energy spinor and 
the negative energy spinor, respectively.
As is well known, negative energy solutions can be interpreted as antiparticles with 
opposite charges from and the same momenta as those of the particles even when their momentum eigenvalue is 
$-p_\mu=-(p_0, -{\bf p})$ \cite{Ryder1}. 
In this paper, we will use the term negative energy spinor because its use is convenient for a one-particle state.

Using the Lorentz covariance of the Dirac equation in Eq. (\ref{eq:Dirac}), one can obtain 
the solutions for a moving particle from the solutions in the particle rest frame by using the Lorentz transformation.
The momentum of particle $p_\mu$ is determined from the momentum of the rest particle 
$k_\mu=(m,0,0,0)$ such as $p_\mu =\left( L_{\bf p} \right)_\mu^{\phantom{\mu}\nu} k_\nu$ 
by using the standard Lorentz boost $L_{\bf p}$.
Then, the spinors $u(p,s)$ and $v(p,s)$ for moving particles can be obtained by  
the spinors $u(k,s)$ and $v(k,s)$ for the rest particle multiplying by $\mathcal{S}(L_{\bf p})$, respectively. 
That is, $u(p,s)=\mathcal{S}(L_{\bf p})u(k,s)$ and $v(p,s)=\mathcal{S}(L_{\bf p})v(k,s)$.
The $\mathcal{S}(L_{\bf p})$ is the Lorentz transformation in the spinor representation 
corresponding to  $L_{\bf p}$:
\begin{eqnarray}
\label{eq:LSR}
\mathcal{S}(L_{\bf p})=\left( \begin{array}{cc} \cosh{\frac{\xi}{2}} & \boldsymbol{\sigma}\cdot \hat{\bf p} \sinh{\frac{\xi}{2}} \\
 \boldsymbol{\sigma}\cdot \hat{\bf p} \sinh{\frac{\xi}{2}} & \cosh{\frac{\xi}{2}} \end{array} \right)
 = \frac{E+m - \gamma^0 \gamma^i p_i}{\sqrt{2m(E+m)}},
\end{eqnarray}
where $\xi$ is the rapidity of the particle and $\hat{\bf p}={\bf p}/\sqrt{{\bf p}^2}$.
 In the particle rest frame, the meaning of the spin index $s$ is clear because 
 the spin operator in the particle rest frame is 
 \begin{eqnarray}
 \label{eq:RestSpin}
 \frac{\boldsymbol{\Sigma}}{2}= \frac{1}{2} \left( \begin{array}{cc}  {\boldsymbol{\sigma}} & 0 \\ 
 0 & {\boldsymbol{\sigma}}  \end{array} \right),
 \end{eqnarray} 
 where $\boldsymbol{\Sigma}$ and $\boldsymbol{\sigma}$ are the three-dimensional vectors 
 $(\Sigma_x, \Sigma_y, \Sigma_z)$ and $(\sigma_x,\sigma_y,\sigma_z)$, respectively.
The spinors $u(k,\pm)$ and $v(k,\pm)$ are eigen-spinors of the 4-dimensional 
Pauli spin matrix $\Sigma_z$ with eigenvalue $\pm 1$. 
 
Note that the spinors $u(p,s)$ and $v(p,s)$ for moving particles have the same spin index $s$ as the rest spinors 
$u(k,s)$ and $v(k,s)$. This is reasonable because Wigner rotation by a single standard Lorentz 
transformation becomes trivial \cite{Wigner}. 
However, the spin operator in the laboratory frame is not clear.
The spin operator $\frac{{\Sigma}_i}{2}$ in Eq. (\ref{eq:RestSpin})  
is $-{W_i}/{2m}$ in the particle rest frame, and  
$W_i$ is the space part of the Pauli-Lubanski vector
\begin{eqnarray}
W_\mu=-\frac{1}{2} \epsilon_{\mu\nu\lambda\delta}J^{\nu \lambda}P^\delta,
\end{eqnarray}
where $J^{\nu \lambda}$ and $P^\delta$ are the angular momentum and the momentum operators, respectively. 
$ \epsilon_{\mu\nu\lambda\delta}$ is a Levi-Civita symbol.
The $W_i$ in the laboratory frame, however, does not satisfy the required spin commutation relations 
$[W_i,W_j]= -i\epsilon_{ijk}W_k/m $ although $W_\mu W^\mu=-m^2 S(S+1)$ is the second Casimir invariant of the 
inhomogeneous Lorentz group (Poincar$\acute{\mbox{e}}$ group) and $S$ is the spin of the particle \cite{Ryder1}.

The proper way to obtain the relativistic spin operator in a fully covariant way is to define the following
spin tensor for the particle rest frame:
\begin{eqnarray}
\Sigma^{\mu \nu}_{0} =\frac{i}{2}[\gamma^\mu, \gamma^\nu].
\end{eqnarray}
In the standard representation, the 3-dimensional spin vector defined by 
$\Sigma_i= \epsilon_{ijk}\Sigma^{jk}_0/2$ becomes the $4\times 4$ Pauli spin $\Sigma_i$, 
where $\epsilon_{ijk}$ is a Levi-Civita symbol.
The spin operator for a moving frame can naturally be defined, by using the Lorentz transformation, from the spin tensor
for the particle rest frame as
\begin{eqnarray}
\Sigma^{\mu\nu}_{\mbox{\scriptsize{R}}} = \mathcal{S}(L_{\bf p}) \Sigma^{\mu \nu}_{0}\mathcal{S}^{-1}(L_{\bf p}),
\end{eqnarray}
where $\mathcal{S}^{-1}(L_{\bf p})$ is the inverse Lorentz transformation of $\mathcal{S}(L_{\bf p})$.
Then the spin operator becomes 
\begin{eqnarray}
\label{eq:RCSpin}
(\Sigma_{\mbox{\scriptsize{R}}} )_i = \frac{\epsilon_{ijk}}{2}\Sigma_{\mbox{\scriptsize{R}}} ^{jk} 
 = \frac{{\Sigma}_i}{2} + \frac{\epsilon_{ijk}p_j (\mbox{\boldmath{$\Sigma$}} \times {\bf p})_k}{2m(E+m)} + i \frac{\gamma_5}{2m} (\mbox{\boldmath{$\Sigma$}}\times {\bf p})_i,
\end{eqnarray}
which is the same as the covariant relativistic spin operator in Ref. 20.
The $\gamma_5=i \gamma^0 \gamma^1 \gamma^2 \gamma^3$ is a $4\times 4$ matrix with off-diagonal terms.
Notice that the eigenvalue equations of the covariant relativistic spin operator in the laboratory frame are
\begin{eqnarray}
(\Sigma_{\mbox{\scriptsize{R}}})_z u(p,\pm) &=& \pm u(p,\pm), \\ \nonumber
 (\Sigma_{\mbox{\scriptsize{R}}})_z v(p,\pm) &=& \pm v(p,\pm).
\end{eqnarray}
These confirm that the spin eigenvalues in the laboratory frame are the same as 
the spin eigenvalues in the particle rest frame. An arbitrary positive energy spin state  
$\psi(k)= a u(k,+) + b u(k,-)$ in the particle rest frame becomes the spin state 
$\psi(p)=\mathcal{S}(L_{\bf p})\psi(k,s)=a u(p,+)+b u(p,-)$ in the moving frame. Then, the expectation value 
of the spin in the particle rest frame $\bar{\psi}(k) \boldsymbol{\Sigma} \psi(k)$ is equal to 
the expectation value of the spin in the moving frame $\bar{\psi}(p) 
\boldsymbol{\Sigma}_{\mbox{\scriptsize{R}}} \psi(p)$, where $\bar{\psi}(p)=\psi(p)^\dagger \gamma_0$ 
and the relativistic invariant normalizations
\begin{eqnarray}
\label{eq:Nor}
\bar{u}(p,s) u(p,s')=-\bar{v}(p,s)v(p,s')=\delta_{ss'}, ~ 
\bar{u}(p,s) v(p,s')= 0
\end{eqnarray}
 are used.
This implies that the measurements of the spin along the same axes by two 
observers, one in the particle rest frame and the other in the laboratory frame, will
give the same spin expectation value.  

It is interesting to compare the covariant relativistic spin operator in Eq. (\ref{eq:RCSpin}) 
with the classical covariant spin in classical electrodynamics.
The classical covariant spin can be defined as ${\bf S}=\gamma_v\hat{\boldsymbol{\mu}}/\alpha$ 
by using the classical covariant magnetic dipole moment \cite{Penfield}
\begin{eqnarray}
\hat{\boldsymbol{\mu}}=\alpha\left[ \frac{\boldsymbol{\sigma}}{2} - 
\frac{{\bf p}({\bf v}\cdot {\boldsymbol{\sigma}})}{2(E+m)}\right],
\end{eqnarray}
 where $\alpha$ is the gyromagnetic ratio. The Lorentz factor $\gamma_v$ is $1/\sqrt{1- {\bf v}^2}$ and 
 ${\bf p}=m \gamma_v {\bf v}$. 
 Notice that the classical covariant spin $\bf S$ can also be derived by using the spin tensor 
 $S_{\mu\nu}=- \epsilon_{\mu\nu\lambda\delta} P^\lambda W^\delta$. 
 One can see the main difference between the covariant relativistic spin 
 operator $\boldsymbol{\Sigma}_{\mbox{\scriptsize{R}}} $ and the classical spin operator 
 ${\bf S}$ is the $\gamma_5$ proportional term.
 That is, without the $\gamma_5$ proportional term, the covariant relativistic spin operator 
  $\boldsymbol{\Sigma}_{\mbox{\scriptsize{R}}} $ becomes ${\bf S} \mathcal{I}$, 
  where $\mathcal{I}$ is the $4\times 4$ identity matrix.
 Therefore, the $\gamma_5$ proportional term is a purely quantum-mechanical term that is manifest 
 in the 4-dimensional spinor representation.\\

{\raggedright\subsection{Construction by Using Commutators of Pauli-Lubanski Vectors}}
The Dirac particles satisfy the inhomogeneous Lorentz group (Poincar$\acute{\mbox{e}}$ group) symmetry. 
There are two Casimir operators in the Poincar$\acute{\mbox{e}}$ group, the first Casimir invariant
$P^\mu P_\mu$ and the second Casimir invariant $W^\mu W_\mu$. 
Thus, it was expected that the covariant relativistic spin operator could be constructed directly 
by using the Pauli-Lubanski vector.
The covariant relativistic spin operator, however, cannot be obtained by using simple linear combinations 
of the components of the Pauli-Lubanski vector. 
The only candidate for the component of the angular momentum vector that is 
a linear function of $W_\mu$ was shown to be \cite{Bogolubov}
\begin{eqnarray}
\frac{1}{M}\left(W_i -  \frac{P_i W^0}{m+E}\right).
\end{eqnarray}
In fact the above equation is the spin operator $\Sigma_i$ in the particle rest frame which is written 
in the laboratory frame by using $W_\mu$.
 
Ryder has given the correct way to construct the spin operator, different from the usual Pauli spin operator $\Sigma_i$, 
by using the Pauli-Lubanski vector \cite{Ryder99}. 
First, the following two antisymmetric tensor operators are defined by using the commutators of the Pauli-Lubanski vectors:
\begin{eqnarray}
W^{\mu \nu} = \frac{1}{m^2} [W^\mu, W^\nu], ~~\tilde{W}^{\mu \nu} = \frac{1}{2} \epsilon^{\mu \nu \rho \delta} W_{\rho \delta}.
\end{eqnarray}
Then, the two tensor operators, 
\begin{eqnarray}
X^{\mu \nu} = -i \left(W^{\mu \nu} + i \tilde{W}^{\mu \nu}\right),  ~
Y^{\mu \nu} = -i \left(W^{\mu \nu} - i \tilde{W}^{\mu \nu}\right), 
\end{eqnarray}
satisfy the commutation relations for the angular momentum. 
Therefore, these two tensor operators can be represented by $2\times 2$ matrices 
which cannot be irreducible representations under the parity. 
The 4-dimensional irreducible spin tensor including parity can be constructed as
\begin{eqnarray}
\left({\Sigma}_{\mbox{\scriptsize{R}}} \right)_{\mu\nu} = \frac{1}{2}\left( 1- \gamma_5 \right) X_{\mu\nu} +  
\frac{1}{2}\left( 1- \gamma_5 \right) Y_{\mu\nu}.
\end{eqnarray}
Then, the spatial component defined by $
(\Sigma_{\mbox{\scriptsize{R}}} )_i = \frac{1}{2} \epsilon_{ijk} (\Sigma_{\mbox{\scriptsize{R}}} )_{jk}$ 
becomes the same as the covariant relativistic spin operator in Eq. (\ref{eq:RCSpin}).\\

{\centering\section{Canonical approach}\label{sec:CAP}}

In this section, we will review the canonical approach given by Foldy and Woutheysen \cite{Foldy}. 
The covariant relativistic spin operator was obtained in the last section; however, it is not clear 
whether the covariant relativistic spin operator is a good quantum observable in the sense 
that good quantum observables must commute with the Hamiltonian of a particle. 
In the canonical approach, the following free Dirac Hamiltonian is used to describe a massive Dirac particle 
in the laboratory frame:
\begin{eqnarray}
\label{eq:DiracH}
\mathcal{H}_{\mbox{\scriptsize{D}}}({\boldsymbol {\mathcal{P}}}) = \gamma^0 m - \gamma^0 \gamma^i \mathcal{P}_i. 
\end{eqnarray}
The $\mathcal{P}_i$ is a momentum operator such that the eigenvalues for a positive energy state and a negative 
energy state are $\pm p_i$, respectively. 
In the particle rest frame, the covariant relativistic spin operator and the free Dirac Hamiltonian 
commute with each other. In the laboratory frame, however, the covariant relativistic spin operator 
$\boldsymbol{\Sigma}_{\mbox{\scriptsize{R}}} $ does not commute with the free Dirac Hamiltonian in Eq. (\ref{eq:DiracH}) 
because the Dirac Hamiltonian in the laboratory frame cannot be obtained simply, by using the Lorentz transformation, 
from the Dirac Hamiltonian in the particle rest frame as a covariant relativistic spin operator. 

Foldy and Woutheysen (FW) found the mean spin operator 
that commutes with the Dirac Hamiltonian for a moving particle
 by using the following canonical transformation \cite{Foldy}:
\begin{eqnarray}
\label{eq:Unitary}
\mathcal{U}({\boldsymbol {\mathcal{P}}})=\frac{E+m- {\gamma}^i{\mathcal{P}_i}}{\sqrt{2E(E+m)}}.
\end{eqnarray}
In the new representation, the Dirac Hamiltonian has a diagonal form such as
 $\tilde{\mathcal{H}}_D({\boldsymbol {\mathcal{P}}})$$=$
 $\mathcal{U}({\boldsymbol {\mathcal{P}}})\mathcal{H}_D({\boldsymbol {\mathcal{P}}}) 
 \mathcal{U}^\dagger({\boldsymbol {\mathcal{P}}})$
 $=$$\gamma^0 E_{\bf p}$, where 
$E_{\bf p}$ represents the operator that gives eigenvalues $\pm E=\pm \sqrt{{\bf p}^2+m}$. 
We call this new representation the FW representation.
In the FW representation, a spinor transforms as 
\begin{eqnarray}
\tilde{\psi}^{\pm}(p)=\sqrt{\frac{m}{E}}\mathcal{U}({\pm \bf p})\psi^{\pm}(p),
\end{eqnarray}
where $\psi^{\pm}(p)$ are the positive energy and the negative energy Dirac spinors in the standard representation 
such that ${\boldsymbol {\mathcal{P}}} \psi^\pm (p) = \pm {\bf p}  \psi^\pm (p)$.
We use the $\tilde{}$ symbol for objects in the FW representation.
Here, the factor $\sqrt{\frac{m}{E}}$ is required because of the different normalizations 
in the canonical representation and in the covariant representation.  
 In the canonical representation, $\tilde{\psi}^\pm(p)^\dagger \tilde{\psi}^\pm(p)=1$ normalizations are used, but 
${\psi}^\pm(p)^\dagger \gamma^0 {\psi^\pm(p)}=\pm 1$ normalizations are used in the covariant representation. 

Notice that the Dirac Hamiltonian of the moving particle in the FW representation has the diagonal form $\gamma^0 E_{\bf p}$.
Therefore, the $4\times 4$ Pauli spin operator $\boldsymbol{\Sigma}$ commutes with
the transformed Hamiltonian $\tilde{\mathcal{H}}_{\mbox{\scriptsize{D}}}({\boldsymbol {\mathcal{P}}})$; 
hence, the spin operator 
$\boldsymbol{\Sigma}$ becomes a good spin observable in the FW representation. 
 The spin operator $\tilde{\boldsymbol{\Sigma}}$ for a moving particle in the FW representation, 
 which is $\boldsymbol{\Sigma}$, 
 has the following positive and negative energy eigen-spinors:  
\begin{eqnarray}
\label{Eq:FWSpinor}
&& \tilde{u}(p,+)=(1,0,0,0)^{\mbox{\scriptsize{T}}} ,~~ 
\tilde{u}(p,-)=(0,1,0,0)^{\mbox{\scriptsize{T}}}, \\ \nonumber
&&\tilde{v}(p,+)=(0,0,1,0)^{\mbox{\scriptsize{T}}},~~ 
\tilde{v}(p,-)=(0,0,0,1)^{\mbox{\scriptsize{T}}},
\end{eqnarray}
where the superscript ${\mbox{\scriptsize{T}}}$ means the transpose of a vector.

The spin operator $\tilde{\boldsymbol{\Sigma}}$ in the FW representation is transformed to   
the FW mean spin operator ${\boldsymbol{\Sigma}}_{\mbox{\scriptsize{FW}}}$ in the standard representation as 
\begin{eqnarray}
\label{eq:FWSpin}
\frac{{\boldsymbol{\Sigma}}_{\mbox{\scriptsize{FW}}}}{2} = \mathcal{U}^\dagger({\boldsymbol {\mathcal{P}}}) 
\frac{\boldsymbol{\tilde{\Sigma}}}{2} \mathcal{U}({\boldsymbol {\mathcal{P}}}) ~
=  \frac{\boldsymbol{\Sigma}}{2} - \frac{i\gamma^0 (\boldsymbol{\Sigma}\times {\boldsymbol {\mathcal{P}}})}{2E} 
- \frac{{\boldsymbol {\mathcal{P}}} \times  (\boldsymbol{\Sigma}\times {\boldsymbol {\mathcal{P}}})}{2E(E+m)}.
\end{eqnarray}
One can easily check that the FW spin operator ${\boldsymbol{\Sigma}}_{\mbox{\scriptsize{FW}}}/2$ commutes with 
the Dirac Hamiltonian $\mathcal{H}_{\mbox{\scriptsize{D}}}({\boldsymbol {\mathcal{P}}})$ by direct calculation. 
This means the FW mean spin operator is a constant of motion and
becomes a good quantum observable in the standard representation. 
As mentioned in Ref. 18, the spin operator ${\boldsymbol{\Sigma}}_{\mbox{\scriptsize{FW}}}$ is nonlocal 
because it depends on the momentum. 
In fact, the eigenvalue of the FW mean spin operator represents the average spin of a rapidly oscillating particle within its 
Compton wavelength. This is the meaning of the name 'FW mean spin operator.' 
The FW mean spin operator has been shown to be the spin operator in the non-relativistic Pauli Hamiltonian \cite{Foldy}. 
This implies that the expectation value measured by experiments in non-relativistic quantum mechanics
is the expectation value of this spin operator. \\

{\centering\section{Equivalence between relativistic spin and FW mean spin}\label{sec:Equ}}

The covariant relativistic spin has been shown by several authors to be equivalent to the Foldy-Woutheysen spin
for the positive energy states \cite{Gursey}.
For the positive energy spinors, the spinor representation $\mathcal{S}(L_{\bf p})$ of the standard Lorentz transformation 
$L_{\bf p}$ in Eq. (\ref{eq:LSR}) can be represented by the unitary operator 
$\mathcal{U}({\boldsymbol {\mathcal{P}}})$ in Eq. (\ref{eq:Unitary})
with the normalization factor $\sqrt{\frac{m}{E}}$, so the equivalence of the two spin operators can 
be given at the operator level. 
The spinor representation of the Lorentz transformation acting on the entire spinor space, however, cannot be a unitary operator because
the Lorentz group has no finite dimensional unitary representation.
At this stage, therefore, it is unclear that the covariant relativistic spin operator is equivalent to 
the FW mean spin operator for negative energy states.
In fact, the two spin operators, the covariant relativistic spin $\boldsymbol{\Sigma}_{\mbox{\scriptsize{R}}}$ 
in Eq. ({\ref{eq:RCSpin}) and the FW mean spin $\boldsymbol{\Sigma}_{\mbox{\scriptsize{FW}}}$ 
in Eq. (\ref{eq:FWSpin}), are different.
This raises a question as to which spin operator is the proper spin observable for a massive Dirac particle.
We will show that they are equivalent in the sense that they give the same 
expectation values. % except the normalization factors.

The equivalence of the two spin operators can be obtained by using the relations between the actions of 
the Lorentz transformation $\mathcal{S}(L_{\bf p})$ and the unitary transformation 
$\mathcal{U}({\boldsymbol {\mathcal{P}}})$ on positive and negative energy spinors.
The unitary operator $\mathcal{U}({{\boldsymbol {\mathcal{P}}}})$ has the momentum operator 
${\boldsymbol {\mathcal{P}}}$, which
gives different eigenvalues for positive energy and negative energy spinors. 
Thus, it is convenient to use the projection operators $\Pi=(m+\gamma^\mu \mathcal{P}_\mu)/2$ such that 
\begin{eqnarray}
{\Pi} u(p,s) &=& \frac{m+\gamma^\mu p_\mu}{2m}u(p,s)=u(p,s),  \\ 
{\Pi}v(p,s) &=& \frac{m-\gamma^\mu p_\mu}{2m}v(p,s)=v(p,s).
\end{eqnarray} 
One can define $\Pi_+=(m+\gamma^\mu p_\mu)/2$ and $\Pi_-=(m-\gamma^\mu p_\mu)/2$, which satisfy 
${\Pi}_+^2 = {\Pi}_+$, ${\Pi}_-^2 ={\Pi}_-$ and $\Pi_- \Pi_+=\Pi_+ \Pi_-=0$. 
That is, $\Pi_+$ and $\Pi_-$ project out positive and negative energy spinors, respectively.
Then, the following relations hold:
\begin{eqnarray}
\mathcal{U}({\bf p}) \Pi_+ &=& \sqrt{\frac{E}{m}}\mathcal{S}^{-1}(L_{\bf p}) {\Pi}_+, ~
\mathcal{U}(-{\bf p}) {\Pi}_- = \sqrt{\frac{E}{m}}\mathcal{S}^{-1}(L_{\bf p}) {\Pi}_-, \\ \nonumber
{\Pi}_+^\dagger \mathcal{U}^\dagger({\bf p}) &=& \sqrt{\frac{E}{m}} {\Pi}_+^\dagger \gamma^0 \mathcal{S}(L_{\bf p}),
~
{\Pi}_-^\dagger \mathcal{U}^\dagger(-{\bf p}) = -\sqrt{\frac{E}{m}}  {\Pi}_-^\dagger \gamma^0 \mathcal{S}(L_{\bf p}).
\end{eqnarray}
Using these relations, we obtain the equivalences
\begin{eqnarray}
\label{eq:EquivalencePo}
\frac{m}{E} \tilde{u}^\dagger(p,s)   \tilde{\boldsymbol{\Sigma}} 
 \tilde{u}(p,s) =  \frac{m}{E} {u}^\dagger(p,s)  \boldsymbol{\Sigma}_{\mbox{\scriptsize{FW}}} 
  {u}(p,s)
=  \bar{u}(p,s) \boldsymbol{\Sigma}_{\mbox{\scriptsize{R}}} u(p,s)
\end{eqnarray}
and 
\begin{eqnarray}
\label{eq:EquivalenceNe}
\frac{m}{E} \tilde{v}^\dagger(p,s) \tilde{\boldsymbol{\Sigma}} \tilde{v}(p,s) = \frac{m}{E} {v}^\dagger(p,s)  \boldsymbol{\Sigma}_{\mbox{\scriptsize{FW}}} 
  {v}(p,s)  
=  -\bar{v}(p,s) \boldsymbol{\Sigma}_{\mbox{\scriptsize{R}}} v(p,s)
\end{eqnarray}
because $\Pi_+ \tilde{u}(p,s)=\tilde{u}(p,s)$ and $\Pi_- \tilde{v}(p,s)=\tilde{v}(p,s)$.
The $-$ sign and $m/E$ are due to the relativistic invariant normalization in Eq. (\ref{eq:Nor}).

Equations (\ref{eq:EquivalencePo}) and (\ref{eq:EquivalenceNe}) guarantee that the expectation values of 
the FW mean spin operator and the covariant relativistic spin operator are the same for the general spinor 
$\psi(p)$, which is a linear combination of the positive and the negative eigen-spinors.
In this sense, the two spin operators are equivalent. As a result, 
the covariant relativistic spin operator can be considered as a good quantum observable 
because it is equivalent to the FW mean spin operator that commutes with the free Dirac Hamiltonian.\\

{\centering\section{Reduced Spin Density Operator}\label{sec:RSD}}

For the study of a reduced spin density operator, the momentum representation of the covariant Dirac equation 
is convenient. 
The solutions of the covariant Dirac equation in the momentum representation 
can be represented as $|p\rangle \otimes u(p,s)$ for positive energy solutions and 
$|-p\rangle \otimes v(p,s)$ for negative energy solutions. 
$|p\rangle$ is the momentum eigenstate; i.e., $\mathcal{P_\mu}|p\rangle=p_\mu|p\rangle$. 
The spin expectation value of a spinor in the laboratory frame 
has been shown to be the same as the spin expectation value of the spinor in the particle rest frame.   
This implies that a reduced spin density matrix can be clearly defined by a 
partial tracing over momentum degrees of freedom.
To make this point more transparent, we will study the reduced spin density matrices in the FW representation because 
 the spin and the momentum degrees of freedoms are decoupled in the FW representation as are those 
 in the particle's rest frame. In the FW representation, the meaning of the partial trace over the momentum is clear.

We consider the following state in the FW representation:
\begin{eqnarray}
\label{eq:IniState}
\tilde{\psi} = \int d^3{\bf p} \tilde{\psi}(p) \otimes |p\rangle
= \int d^3{\bf p} \sum_{\delta} \left[a_\delta({\bf p}) \tilde{u}(p,\delta) +
b_\delta({\bf p}) \tilde{v}(p,\delta) \right]\otimes |p\rangle.
\end{eqnarray}
The normalization $\tilde{\psi}^\dagger \tilde{\psi}$$=$$1$ requires $\int d^3 {\bf p}$$ \sum_{\delta} $$
[ a_\delta({\bf p}) a^*_\delta({\bf p})$$ + $$b_\delta({\bf p}) b_\delta^*({\bf p})]$$=$$1$.
For the density matrix $\tilde{\rho}= \tilde{\psi}\tilde{\psi}^\dagger$, the reduced spin density matrix is well defined by a partial trace over momentum as
\begin{eqnarray}
\label{eq:RedSpinDensity}
\tilde{\rho}_s &=& \mbox{Tr}_{\bf p} \tilde{\rho} = \int d^3 {\bf p} \tilde{\psi}^\dagger(p) \tilde{\psi}(p)
\\ \nonumber
&=& \int d^3{\bf p}\sum_{\lambda, \lambda'} \left[ a_\lambda({\bf p}) a^*_{\lambda'}({\bf p})
 \tilde{u}(p,\lambda) \tilde{u}^\dagger (p,\lambda') 
  + a_\lambda({\bf p}) b^*_{\lambda'}({\bf p}) \tilde{u}(p,\lambda) 
 \tilde{v}^\dagger(p,\lambda') \right. \\ \nonumber
  && \phantom{\int d^3{\bf p}} \left. + b_\lambda({\bf p}) a^*_{\lambda'}({\bf p}) \tilde{v}(p,\lambda) 
  \tilde{u}^\dagger(p,\lambda') 
  + b_\lambda({\bf p}) b^*_{\lambda'}({\bf p}) \tilde{v}(p,\lambda) \tilde{v}^\dagger(p,\lambda') \right].
\end{eqnarray}
The reduced spin density matrix becomes the same as the non-relativistic spin density matrix 
for a positive energy particle. 

The relativistic effects on the reduced spin density matrix of a moving particle is represented by 
the transformation of the reduced spin density matrix under a general Lorentz transformation  
The transformation of the reduced spin density matrix is given by the transformation matrix of the spinor 
under the Lorentz transformation.
We have obtained the transformation matrix for the positive energy spinors in a previous work \cite{Choi}.
In this section, the complete transformation matrix for the whole of spinor space is obtained. 
The complete transformation matrix in the FW representation will be shown to be equivalent to the complete 
transformation matrix in the covariant representation.

Let us assume that the two observers $\mathcal{O}$ and $\mathcal{O}'$ are related by an arbitrary Lorentz transformation $\Lambda$. Lorentz transformations do not change the sign of the energy for a free particle, 
so we must deal with the positive and the negative energy states separately.
The transformation matrix $\mathcal{T}^{(\Lambda, {\bf p})}$ in the covariant representation and 
the transformation matrix $\tilde{\mathcal{T}}^{(\Lambda, {\bf p})}$ in the FW representation 
for the positive energy spinors have the following relations:
\begin{eqnarray}
\mathcal{S}(\Lambda)u( p,\lambda) &=& \mathcal{S}(\Lambda) \mathcal{S}(L_{\bf p}) u(k,\lambda) 
   =   \sum_{\lambda'} {\mathcal{T}}^{(\Lambda, {\bf p})}_{\lambda' \lambda} 
   \mathcal{S}(L_{\Lambda \bf p}) u(k,\lambda')\\ \nonumber
   &=& \sqrt{\frac{E}{m}}\mathcal{S}(\Lambda) \mathcal{U}^\dagger({\boldsymbol {\mathcal{P}}}) 
   \tilde{u}(p,\lambda)
   =   \sqrt{\frac{E}{m}}\sum_{\lambda'} \tilde{\mathcal{T}}^{(\Lambda, {\bf p})}_{\lambda' \lambda} 
   \mathcal{U}^\dagger({\boldsymbol {\mathcal{P}}} ) \tilde{u}(\Lambda p,\lambda') ,
\end{eqnarray}
where ${\Lambda}{\bf p}$ is the space part of the four momentum 
$(\Lambda p)_\mu$$=$$ \Lambda_\mu^{\phantom{\mu}\nu} p_\nu$.
This equation shows that the transformation matrix $\mathcal{T} ^{(\Lambda, {\bf p})}_{\lambda' \lambda}$ $=$ 
$ u (k, \lambda') \mathcal{S}^{-1}(L_{\Lambda {\bf p}})  \mathcal{S}(\Lambda) \mathcal{S}(L_{\bf p})  u(k,\lambda) $ 
in the covariant representation is equivalent to the transformation matrix 
$\tilde{\mathcal{T}} ^{(\Lambda, {\bf p})}_{\lambda' \lambda}$ $=$ 
$ \tilde{u}(\Lambda p, \lambda') \mathcal{U}({\boldsymbol {\mathcal{P}}} ) \mathcal{S}(\Lambda) 
\mathcal{U}^\dagger({\boldsymbol {\mathcal{P}}}) \tilde{u}(p,\lambda)$ in the FW representation
because the forms of $u(k,\lambda) $ and $\tilde{u}(p,\lambda)$ are the same.
The transformation matrices for the negative energy spinor is obtained in a similar manner. As a result, 
the complete transformation matrix can effectively be written in the following block diagonal form: 
\begin{eqnarray}
\label{Eq:TransM}
\tilde{\mathcal{T}}^{\Lambda, {\bf p}}=\left( \begin{array}{cccc} A& B& 0& 0 \\
-B^* & A & 0 &0 \\
0 & 0 & A & B \\
0 & 0 & -B^* & A \end{array} \right),
\end{eqnarray}
where $|A|^2+|B|^2=1$ and $A^* B - AB=0$.
This transformation matrix shows that the irreducible representation for the Lorentz transformation $\Lambda$
is the $2\times 2$ unitary matrix $\left( \begin{array}{cc} A& B\\-B^* & A \end{array} \right)$.
This two-dimensional matrix corresponds to the two-dimensional unitary representation of the Wigner rotation.
This means the FW representation and the covariant representation are two equivalent 
representations for Wigner's little group.

Now, we will consider the change in the reduced spin density matrix $\tilde{\rho}_s$ 
under the Lorentz transformation $S(\Lambda)$.
The transformation of the spinor $\tilde{\psi}(p)$ is obtained as 
\begin{eqnarray}
\tilde{\psi}(\Lambda p)  = \mathcal{U}({\boldsymbol {\mathcal{P}}}) \mathcal{S}(\Lambda)
\mathcal{U}^\dagger({\boldsymbol {\mathcal{P}}})
  \tilde{\psi}(p)  
  = \sum_{\lambda', \lambda} \tilde{\mathcal{T}}^{\Lambda, {\bf p}}_{\lambda' \lambda} 
   \left( a_{\lambda}({\bf p}) \tilde{u}(\Lambda p,\lambda')+ 
   b_{\lambda}({\bf p}) \tilde{v}(\Lambda p,\lambda') \right).
\end{eqnarray}
The reduced spin density matrix for the observer $\mathcal{O}'$, defined by 
\begin{eqnarray} 
\label{eq:TRDen}
\tilde{\rho'}_s = \int d^3 {\bf p} \tilde{\psi}(\Lambda p)  \tilde{\psi}(\Lambda p),
\end{eqnarray}
shows the dependence of the rotation of the spin not only on the Lorentz transformation $\Lambda$ 
but also on the momentum of the particle.
This momentum-dependent rotation changes the entanglement between the spin and the momentum degrees of freedom. 
As a result, the spin entropy, which describes the mixedness of the reduced spin density matrix, changes.

The reduced spin density matrix for the general state $\tilde{\psi}$ in Eq. (\ref{eq:IniState}) 
is too complicated to study the essential feature of the spin density matrix under the general Lorentz transformation.
Therefore, we will consider a simple example that shows the nontrivial effects under the Lorentz transformation.
We consider the following two initial positive energy states in the FW representation:
\begin{eqnarray}
\label{eq:States}
\tilde{\psi}_1 &=& \frac{1}{\sqrt{2 \delta({\bf 0}) }} \left( \tilde{u}( p,+)\otimes |p\rangle 
+ \tilde{u}( p_\perp,+) \otimes |p_\perp\rangle \right), \\ 
\tilde{\psi}_2 &=&  \frac{1}{\sqrt{2 } \delta({\bf 0})} \left(\tilde{u}( p,+)  \otimes |p\rangle
+ \tilde{u}( p_\perp,-)  \otimes |p_\perp\rangle\right),
\end{eqnarray}
where 
the four momentum $p^\mu=(E,{\bf p})$ and $p_\perp^\mu=(E,{\bf p}_\perp)$. 
$1/\sqrt{2 \delta({\bf 0})}$ is included due to the normalization $\tilde{\psi}^\dagger_1 \tilde{\psi}_1$$
=$$\tilde{\psi}^\dagger_2 \tilde{\psi}_2=1$.
The space parts of the momentum are written in spherical coordinate as ${\bf p}$ $=$
$\{ p \sin{ \theta}  \cos{ \phi },p \sin {\theta } \sin {\phi},p \cos {\theta }\} $ and 
${\bf p}_\perp$$=$$(E, p \cos{\theta} \cos{\phi},p \cos{\theta}\sin{\phi}, -p \sin{\theta})$. 
$ {\bf p}_\perp$ is perpendicular to ${\bf p}$ and corresponds to the replacement of 
$\theta$ by $\theta+\pi/2$. 
$\theta$ is the polar angle from the positive $z$-axis, and $\phi$ is the azimuthal angle in the $xy$ plane from
the $x$-axis. The action of a rotation of an observer is trivial because the rotation does not change 
the entanglement between the momentum degrees of freedom and the spin degrees of freedom.
Therefore, we consider the Lorentz boost $\Lambda _{\xi }$
in the $z$ direction with rapidity $\xi$. This does not reduce the generality because 
the general momentum vector $p^\mu$ is considered. 
The reduced spin density matrices can be described by $2\times 2$ matrices 
because the positive energy spinors in the FW representation can be represented by two-component vectors. 
The reduced spin density matrices $\tilde{\rho}_{1s}$ and $\tilde{\rho}_{2s}$  are obtained by tracing over the momentum for
the density matrices $\tilde{\rho}_1=\tilde{\psi}_1 \tilde{\psi}^\dagger_1$ and 
$\tilde{\rho}_2= \tilde{\psi}_2 \tilde{\psi}^\dagger_2$, respectively. 
The $\tilde{\rho}_{1s}$ becomes the pure state $\tilde{u}(p,+)\tilde{u}^\dagger(p, +)$, 
and the $\tilde{\rho}_{2s}$ becomes the complete mixed state $[\tilde{u}(p,+) \tilde{u}^\dagger(p,+)$$+$
$\tilde{u}(p_\perp,-)\tilde{u}^\dagger(p_\perp,-)]/2$. 
Note that the two-dimensional representations of the spinors $\tilde{u}(p,\pm)$ in the FW representation are described by 
$|\pm\rangle $, which are the eigenstates of the usual Pauli matrix $\sigma_z$ with eigenvalues 
$\pm 1$ such that $\sigma_z|\pm\rangle=\pm|\pm\rangle$.

The transformed spin density matrices  become
\begin{eqnarray}
\label{eq:TrDen1}
\tilde{\rho}'_{1s} &=&\frac{1}{2}  \tilde{\mathcal{T}}^{(\Lambda, {\bf p})} |+\rangle \langle 
+ |\left( \tilde{\mathcal{T}}^{(\Lambda, {\bf p})} \right)^\dagger  
+  \frac{1}{2}\tilde{\mathcal{T}}^{(\Lambda, {\bf p}_\perp)} |+\rangle \langle+|  
\left( \tilde{\mathcal{T}}^{(\Lambda, {\bf p}_\perp)} \right)^\dagger\\ \nonumber
&=& \frac{1}{2} \left( \begin{array}{cc} a_1^2 + a_2^2 & -(a_1 b_1 + a_2 b_2)e^{-i\phi} \\
 -(a_1 b_1 + a_2 b_2)e^{i\phi} & b_1^2 + b_2^2 \end{array} \right)
\end{eqnarray}
and
\begin{eqnarray}
\label{eq:TrDen2}
\tilde{\rho}'_{2s} &=&\frac{1}{2}  \tilde{\mathcal{T}}^{(\Lambda, {\bf p})} |+\rangle \langle  
+ |\left( \tilde{\mathcal{T}}^{(\Lambda, {\bf p})} \right)^\dagger  
+  \frac{1}{2}\tilde{\mathcal{T}}^{(\Lambda, {\bf p}_\perp)} |-\rangle \langle-|  
\left( \tilde{\mathcal{T}}^{(\Lambda, {\bf p}_\perp)} \right)^\dagger\\ \nonumber
&=& \frac{1}{2} \left( \begin{array}{cc} a_1^2 + b_2^2 & (a_2 b_2 - a_1 b_1)e^{-i\phi} \\
 (a_2 b_2 - a_1 b_1)e^{i\phi} & b_1^2 + a_2^2 \end{array} \right).
\end{eqnarray}  
The parameters $a_1$, $a_2$, $b_1$ and $b_2$ are
\begin{eqnarray}
a_1 &=& \sqrt{\frac{m+E}{m+E'}} \left[\cosh \frac{\xi }{2}+\frac{p \cos \theta }{m+E} \sinh \frac{\xi }{2}\right], \\ \nonumber
b_1 &=& \frac{p \sin \theta }{\sqrt{(m+E) \left(m+E'\right)}} \sinh \frac{\xi }{2}, \\
a_2 &=& \sqrt{\frac{m+E}{m+E''}} \left[\cosh \frac{\xi }{2}+\frac{-p \sin \theta }{m+E} \sinh \frac{\xi }{2}\right], \\
b_2 &=&  \frac{p \cos \theta }{\sqrt{(m+E) \left(m+E''\right)}} \sinh \frac{\xi }{2},
\end{eqnarray}
where $E'$$=$$E \cosh \xi $$+$$ p \cos\theta \sinh \xi$ and $E''$$=$$E \cosh \xi $$-$$ p \sin\theta \sinh \xi$.

\begin{figure}
\centering
\includegraphics[]{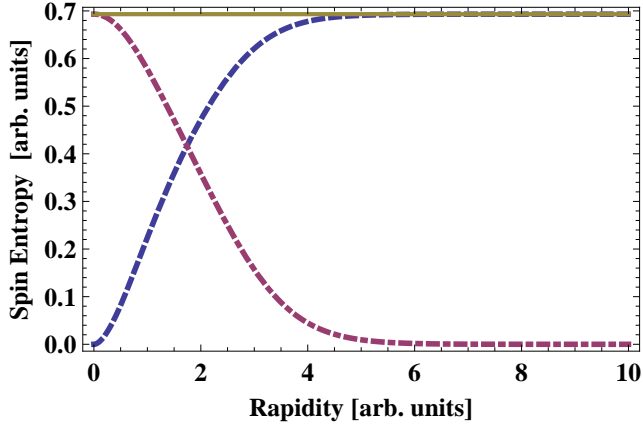}
\caption{(Color on-line) This figure shows the change of entropies for the spin density matrices 
$\tilde{\rho}'_{1s}$ (blue dashed line) and $\tilde{\rho}'_{2s}$ (red dot-dashed line) as functions of 
rapidity with $\sqrt{{\bf p}^2}=10$ and $m=1$ for $\theta=0.54 \pi$. 
The brown real line represents $S=\ln{2}$ corresponding to the completely mixed state.}
\label{fig:1}
\end{figure}

\begin{figure}
\centering
\includegraphics[]{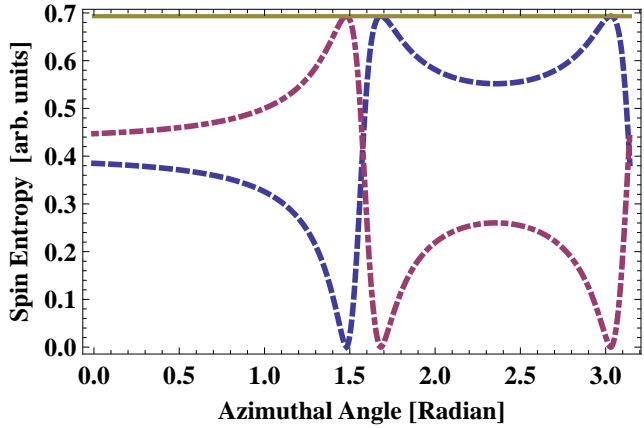}
\caption{(Color on-line) This figure shows the behavior of the spin entropies as functions of the polar angle
$\theta$ for $\tilde{\rho}'_{1s}$ (blue dashed line) and $\tilde{\rho}'_{2s}$ (red dot-dashed line) with
 $\sqrt{{\bf p}^2}=10$ and $m=1$ for $\xi=10$. The brown real line represents $S=\ln{2}$ corresponding to 
 the completely mixed state.}
\label{fig:2}
\end{figure}

We will study the change in the purification of the spin by using the entropy of the spin density matrix.
The entropy for the state $\rho_s$ is defined as \cite{Peres3}
\begin{eqnarray}
S= - \mbox{Tr} \left( \rho_s \ln \rho_s \right)  = - \sum_j \lambda_j \ln \lambda_j,
\end{eqnarray}
where $\lambda_j$ is an eigenvalue of $\rho_s$.
The spin density matrices transform nontrivially under a Lorentz transformation as shown 
in Eq. (\ref{eq:TrDen1}) and Eq. (\ref{eq:TrDen2}), so 
the spin entropies will also change nontrivially. 
The results are shown in Fig. \ref{fig:1} and Fig. \ref{fig:2}. For explicit calculations, we have set 
the momentum $\sqrt{{\bf p}^2}=10$ and the rest mass $m=1$. 
Fig \ref{fig:1} shows the dependence on the rapidity $\xi$ for $\theta=0.54 \pi$. 
The magnitude of the observer's velocity 
becomes $0.9999999959c$ for $\xi=10$. The changes of spin entropies with increasing rapidity in this figure show that  
the pure state can be changed to the completely mixed state and vice versa. 
Fig \ref{fig:2} shows the polar angle dependence of the spin entropies for the rapidity $\xi=10$. 
The spin entropies in this figure show that depending on the polar angle, 
the Lorentz transformation changes the mixedness of the spin over the whole range.

Because the transformation matrices $\mathcal{T}^{(\Lambda, {\bf p})}$ and 
$\tilde{\mathcal{T}}^{(\Lambda, {\bf p})}$ in the covariant representation and in the FW representation
are equivalent, the behaviors of the reduced spin density matrix in the two representations are the same. 
The difference in the two representations are the normalizations. In the covariant representation, 
the density matrix for the observer $\mathcal{O}$ is defined as $\rho=\psi \psi^\dagger \gamma^0$
to guarantee the Lorentz invariance of the normalization.
The trace over momentum is also represented by the Lorentz invariant measure $d^3 {\bf p}/2E$.\\

{\centering\section{Summary}\label{sec:SUM}}

In summary, we have shown the equivalence between the covariant relativistic spin and the FW mean spin of
a Dirac particle. Based on the equivalence, the covariant relativistic spin operator is clearly 
a good spin operator in the covariant representation. 
As a result, the spin index of the Dirac spinor can be understood to represent the 
spin eigenvalues of the moving particle. 
The covariant relativistic spin is shown to have a pure quantum contribution, which cannot be given by 
the classical spin. 
In the FW representation, the Dirac Hamiltonian for a moving particle assumes a diagonal form of
the Hamiltonian in the particle rest frame. 
This fact makes dealing with the momentum and the spin degrees of freedom separately easy.

We have studied the relativistic effects on the spin state in the FW representation.
The spin state can be defined by tracing over the momentum degrees of freedom for the complete density matrix.
The trace over the momentum is obtained by integrating over the momentum, which was considered ambiguous
in the Dirac spinor because of momentum-dependent components.
 This ambiguity can be cleared by considering the problem in the FW representation, which has been shown to be 
 equivalent to the covariant representation.
 The spin entropy, which describes the purity of the spin, changes under the Lorentz transformation.
 The pure spin state can become a totally mixed spin state and vice versa under the Lorentz transformation.
Therefore, the entropy of the spin is neither a Lorentz invariant nor covariant. \\

{\centering\section*{ACKNOWLEDGMENTS}}

The authors are grateful for helpful discussions with Prof. Jaewan Kim at Korea Institute for Advanced Study.
This work was supported by a National Research Foundation of Korea grant funded by the Korean
Government (2011-0005740).


\begin{references}
\bibitem{ChoiNote} Note that the classical magnetic dipole moment in Eq. (11) is the same as the magnetic dipole 
moment in Eq. (6) of Ref. 16.
\end{references}
\end{document}